\documentclass[aps,prc,twocolumn,nofootinbib,preprintnumbers,amsmath,amssymb]{revtex4}

\usepackage{graphicx}
\usepackage{dcolumn}
\usepackage{bm}
\usepackage{epsfig}
\usepackage{longtable}
\usepackage{color}

\def\bea {\begin{eqnarray}}
\def\eea {\end{eqnarray}}
\def\be {\begin{equation}}
\def\ee {\end{equation}}
\def\ben{\begin{enumerate}}
\def\een{\end{enumerate}}
\def\bi{\begin{itemize}}
\def\ei{\end{itemize}}
\def\ie{{\it i.e.}\ }

\def\eg{{\it e.g.}\ }

\def\F{{\cal F}}
\def\GV{G_{\mbox{\tiny V}}}

\def\DRV{\Delta_{\mbox{\tiny R}}^{\mbox{\tiny V}}}

\begin{document}

\preprint{}

\title{
Precise half-life measurement of the superallowed emitter $^{30}$S
}
\author{V.E. Iacob}
\email{iacob@comp.tamu.edu}

\author{J.C. Hardy}
\email{hardy@comp.tamu.edu}

\author{L. Chen}

\author{V. Horvat}

\author{M. Bencomo}

\author{N. Nica}

\author{H.I. Park}

\author{B.T. Roeder}

\author{A. Saastamoinen}

\affiliation{ Cyclotron Institute, Texas A\&M University, College Station, Texas 77843, USA}
\homepage{http://cyclotron.tamu.edu/}

\date{\today}

\begin{abstract}
We have measured the half-life of $^{30}$S, the parent of a superallowed $0^+$$\rightarrow 0^+$ $\beta$ transition, to high precision using very pure sources and
a 4$\pi$ proportional gas counter to detect the decay positrons.  Our result for the half-life is 1.17992(34)\,s.  As a byproduct of this measurement, we determined
the half-life of its daughter, $^{30}$P, to be 2.501(2)\,min.
 
\end{abstract}

\maketitle

\section{\label{sec:introd} INTRODUCTION}

Currently, the superallowed $\beta$-decay branch from $^{30}$S is not one of the fourteen $0^+$$\rightarrow 0^+$ transitions that have been measured to $\pm$0.1\%
precision and collectively used to determine, $V_{ud}$, the up-down element of the Cabibbo-Kobayashi-Maskawa (CKM) quark-mixing matrix. The most recent survey of
world data \cite{HT15} gives $ft$ = 3005(41)\,s for the $^{30}$S superallowed transition, a precision of $\pm$1.4\%, which is more than a factor of 10 too large for
it to contribute to any fundamental tests of the weak interaction or of isospin symmetry breaking.  There are good reasons to improve this situation though,
and we report here a first step in doing so.

The superallowed $0^+$$\rightarrow 0^+$ $\beta$ transitions between $T=1$ analog states are interesting in general because they depend uniquely on the vector part
of the weak interaction. Their experimental $ft$ values are related to the vector coupling constant $\GV $, which, according to conservation of the vector current
(CVC), should be the same for all such transitions.  In practice, several small ($\sim$1\%) correction terms intervene, so these can be combined with
the $ft$ value to obtain a ``corrected" $\F t$ value \cite{HT15}:
\be
\F t \equiv ft(1+\delta_R^{\prime}) (1 + \delta_{NS} - \delta_C)
= \frac{K}{2 \GV^2 (1 + \DRV )}
\label{Ftdef}
\ee
where $K/(\hbar c )^6 = 2 \pi^3 \hbar \ln 2 / (m_e c^2)^5 = 8120.2776(9) \times 10^{-10}$ GeV$^{-4}$s,  $\delta_C$ is the isospin-symmetry-breaking correction and
$\DRV$ is the transition-independent part of the radiative correction.  The terms $\delta_R^{\prime}$ and $\delta_{NS}$ comprise the transition-dependent part of
the radiative correction, the former being a function only of the positron's energy and the $Z$ of the daughter nucleus, while the latter, like $\delta_C$, depends
in its evaluation on the details of nuclear structure.  This $\F t$ value, which combines the experimental $ft$ together with small calculated corrections, is
inversely proportional to $\GV^2$ and thus should be statistically constant from one transition to another.  The precisely determined $\F t$ values from fourteen
superallowed transitions satisfy this condition.  Thus their average can be used to derive the value of $V_{ud}$ \cite{HT15}.  At the present time, this is the most
precise result for $V_{ud}$ and it enables the most demanding available test of the unitarity of the CKM matrix, a test which is passed within $\pm$0.06\%.

Since any deviation from unitarity would be a signal of new physics, there is strong motivation to further improve the unitarity test by reducing the uncertainty
attached to $V_{ud}$.  The largest contributor to its uncertainty is the calculated radiative correction $\DRV$, but the nuclear-structure-dependent corrections
($\delta_C$ - $\delta_{NS}$) are the second largest \cite{HT15} and, unlike $\DRV$, they can be tested by experiment. By requiring that a set of structure-dependent
corrections must produce a statistically consistent set of $\F t$ values or be rejected, we can restrict the number of acceptable sets, and thus reduce the
uncertainties attached to ($\delta_C$ - $\delta_{NS}$) \cite{TH10}.

The $^{30}$S case is a particularly interesting one because its nuclear-structure-dependent correction as calculated with the currently favored ``Shell-model,
Woods-Saxon'' (SM-WS) model \cite{HT15,TH08} is ($\delta_C$ - $\delta_{NS}$) = 1.040(32)\%.  That makes it one of the two largest correction terms among all the
superallowed transitions with $A$$\leq$54, the cases for which the shell model can be considered rather reliable.  If the measured $ft$ value for a transition
with such a relatively large correction were to yield an $\F t$ value that is consistent with the other well-known cases, it would serve to verify the
calculations' reliability for all those cases that have smaller corrections.

The $ft$ value that characterizes any $\beta$ transition depends on three measured quantities: the total transition energy $Q_{EC}$, the half-life $t_{1/2}$ of
the parent state, and the branching ratio $R$ for the particular transition of interest.  For the $^{30}$S superallowed transition, only the $Q_{EC}$ value can
be considered known with sufficient precision.  Its contribution to the $ft$-value uncertainty is merely $\pm$0.03\% \cite{HT15}. The large uncertainty
currently assigned to the $^{30}$S $ft$ value is mostly attributable to the branching ratio of the superallowed transition, which is based on a single 1963
measurement \cite{Fr63}.  Though the half-life is known much more precisely, it is still not adequate since its world-average value \cite{HT15} is quoted to
$\pm$0.14\% and is based on only two measurements, one quoted to $\pm$0.41\% \cite{Wi80} and the other to $\pm$0.14\% \cite{So11}.

We have chosen to begin with a measurement of the half-life of $^{30}$S.  Quite apart from its ultimate benefit in contributing to a usefully precise $ft$ value, 
the measurement also offers an excellent opportunity for us to verify one of the techniques we have used in previous half-life measurements. Unlike most $T_Z$\,=\,-1
superallowed $\beta$ emitters, $^{30}$S does not feed a second $0^+$$\rightarrow 0^+$ $\beta$ transition from its daughter, $^{30}$P.  Instead, the $0^+$, $T$\,=\,1
state populated in $^{30}$P decays electromagnetically to the $1^+$, $T$\,=\,0 ground state, which proceeds by ordinary allowed $\beta$ decay to $^{30}$Si with a
half-life of 2.498(4) min \cite{Ba10, En90}.  This leads to a very clean separation between the $^{30}$S half-life of 1.18 s and that of its daughter, which is
more than a factor of 100 longer.  In the $T_Z$\,=\,-1 cases we have measured before \cite{Ia06, Ia10, Pa11}, the parent and daughter half-lives only differ
by a factor of $\sim$2 and, because we detect the positrons from both decays together in the same detector, we must use the parent-daughter linkage as input to
the fit in order to extract the parent half-life with any precision.  This requires us to know the time-dependence of the source deposit rate and also to incorporate
the subtle difference between the parent and daughter detection efficiencies for positrons \cite{Ia10}; together, these effects can introduce systematic uncertainties.
In the case of $^{30}$S, the large difference between parent and daughter half-lives makes it possible to extract a result for $^{30}$S by treating the two decays as
independent components and to compare that half-life result to the one obtained when the linkage between parent and daughter is enforced.  This is our first
opportunity to make such a comparison.

\section{\label{exp} Experiment}
\subsection{\label{overview} Overview}

We produced a $^{30}$S radioactive beam via the inverse-kinematics reaction $^1$H($^{31}$P, 2$n$)$^{30}$S, initiated by a 30$A$-MeV, $\sim$250-nA beam of $^{31}$P
from the K500 superconducting cyclotron at Texas A\&M University.  The target was hydrogen gas at a pressure of 2 atm, contained in a cell that was kept at
liquid-nitrogen temperature to increase the gas density.  The fully-stripped reaction products exiting the target cell entered the Momentum Achromat Recoil Spectrometer
(MARS) \cite{Tr89} where they were separated according to their charge-to-mass ratio $q/m$, with $^{30}$S being selected in the focal plane.  After passing through the
extraction slits, the $^{30}$S beam exited the vacuum system through a 51-$\mu$m-thick Kapton window, passed through a 0.3-mm-thick BC404 scintillator, where the
ions were counted, and then through a stack of aluminum degraders, finally stopping in the 76-$\mu$m-thick aluminized Mylar tape of a fast tape-transport system.

After the radioactive sample had been collected for a time interval approximately equal to the $^{30}$S half-life, the beam was turned off and the tape-transport system
moved the sample rapidly to a well-shielded location 90 cm away, stopping it at the center of a 4$\pi$ proportional gas counter.  Beginning 220 ms after the beam
was turned off, the decay positrons were recorded for 20 half-lives (24\,s).  The beam was then turned on again and the cycle repeated.  These collect-move-detect
cycles were controlled by the tape-transport system's precision clock, and the timing of the generated control signals was continuously monitored on-line by our
data-acquisition system.  The cycles were repeated until the desired counting statistics had been
achieved.  For this experiment we accumulated data from nearly 6,000 cycles divided into 33 separate runs, which yielded a total of 1.4$\times 10^8$ beta counts. 

\subsection{\label{purity} Source purity}

\begin{figure}[t]
\centering
\includegraphics[width=\columnwidth]{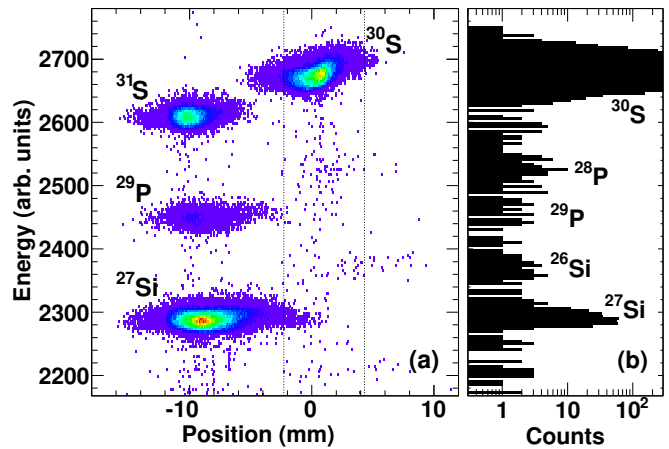}
\caption{\label{fig1} (Color online) Deposited energy versus position as recorded in the PSSD at the MARS focal plane.  This result was obtained after the spectrometer
had been tuned for $^{30}$S, but before the extraction slits had been narrowed. The vertical lines show the final position of the extraction slits, 6.6 mm apart, which we used
during these measurements.  The projection on the right is of the counts between the slits. From such spectra recorded periodically during our experiment, we determined
that the extracted $^{30}$S beam included a 0.37(8)\% contribution from $^{27}$Si and 0.050(15)\% from $^{29}$P.}
\end{figure}
  
Before beginning the measurements, we inserted a 1-mm-thick 16-strip position-sensitive silicon detector (PSSD) into the focal plane of MARS while using a
low-current primary beam. This detector was used first for the identification of secondary reaction products, and then to monitor the selection and focus of
$^{30}$S in the center of the beam line. As shown in Fig.~\ref{fig1}, in addition to $^{30}$S, there were four reaction products, $^{26}$Si ($t_{1/2}$\,=\,2.255\,s),
$^{27}$Si ($t_{1/2}$\,=\,4.135\,s), $^{28}$P ($t_{1/2}$\,=\,270.3\,ms) and $^{29}$P ($t_{1/2}$\,=\,4.140\,s), that appeared between the extraction slits as
weak contaminants in the extracted $^{30}$S beam.  On the one hand, $^{26}$Si and $^{28}$P were not problematic since both passed completely through the tape
without contaminating the collected sample.  On the other hand, $^{27}$Si and $^{29}$P, having a similar range to $^{30}$S, were potentially a concern.  With
the focal-plane acceptance slits of MARS set to a width of 6.6 mm, the total extracted beam contained 0.37(8)\% of $^{27}$Si ions and 0.050(15)\% of $^{29}$P.
The composition of the beam exiting MARS was checked on a daily basis during our half-life measurement: On each occasion we reinserted the PSSD at the MARS focal
plane and acquired a spectrum equivalent to the one shown in Fig.~\ref{fig1}. There were some changes observed from time to time in the extracted beam composition;
these are reflected in the uncertainties we quote on the impurity percentages.

\begin{figure}[t]
\centering
\includegraphics[width=\columnwidth]{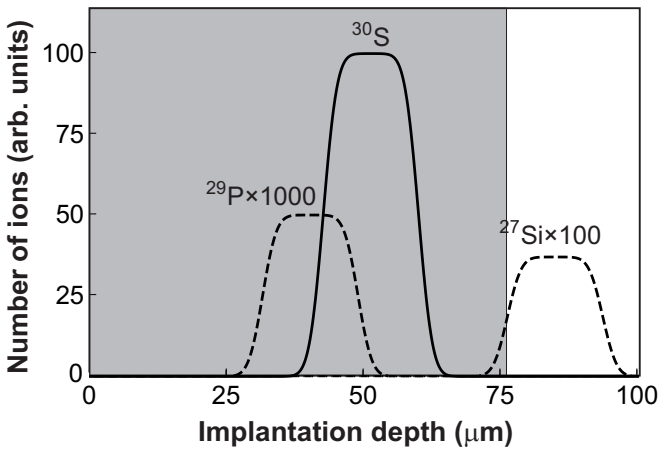}
\caption{\label{fig2} The implantation profiles of $^{30}$S (solid line) and the two contaminant activities, $^{29}$P and $^{27}$Si (dashed lines).  The beam enters
from the left.  The shaded region shows the actual thickness of Mylar collection tape. Our collected source contains only those ions that are stopped inside the tape.}
\end{figure}

Although the amount of $^{27}$Si contaminant from MARS, at 0.37(8)\%, is already quite small, it is large enough to cause concern for a half-life measurement aiming
for sub-0.1\% precision, so we adjusted the thickness of aluminum degraders to minimize the amount of $^{27}$Si that stopped in the tape. This was accomplished as
follows:

Before the main measurements began, we measured the amount of $^{30}$S activity retained in the tape as a function of degrader thickness, spanning the full range from
a minimum, at which all the $^{30}$S passed entirely through the tape, to a maximum, at which no $^{30}$S reached the tape at all.  This yielded a detailed
picture of the depth distribution of the implanted ions, valid not only for $^{30}$S but also for other ions of similar range.  It also produced an experimental value
for the degrader thickness required to center the deposit at the desired depth in the tape.  As in our previous published measurements, the value obtained was very close to that predicted
by the SRIM code \cite{SRIM}, giving confidence that the SRIM calculations can be relied on to determine the spatial distributions of impurities relative to that
of $^{30}$S.  

Figure \ref{fig2} presents the results of calculations based both on the SRIM code and on our measured depth distribution.  Note that they correspond to the thickness
of degraders we actually used in our half-life measurement, which placed the $^{30}$S activity towards the back of the tape.  The solid line in the figure represents a
simplified functional form for the implantation-depth distribution of $^{30}$S; it is consistent with the results of our scan of $^{30}$S activity versus degrader
thickness and with the known momentum spread as set by the momentum-selection slits in MARS ($\Delta p/p$ = 0.86\%).  We have then reproduced the semi-empirical $^{30}$S shape
as dashed lines to represent the distributions of $^{29}$P and $^{27}$Si at their calculated locations and with their measured relative intensities. 

Based on the results illustrated in Fig.\,\ref{fig2}, we determined that only approximately 4\% of the $^{27}$Si impurity observed in the MARS focal plane was actually
retained in the tape with our particular choice of degrader thickness.  This means that its effective contribution to the measurement is 0.015(15)\% relative to that
of $^{30}$S.  The contribution of $^{29}$P, being obviously unaffected by the degrader choice, is the same in the tape as it was in the MARS focal plane.  Coincidentally,
$^{29}$P and $^{27}$Si have almost identical half-lives so their effects on the measurement are equivalent.  Their relative intensities in the tape sum to 0.065(21)\%,
a manageably small amount. 

\subsection{\label{4pi} 4$\pi$ gas counter} 

Our 4$\pi$ gas counter consists of two separate gas cells that, when assembled, leave a very narrow slot between them, through which the aluminized Mylar transport tape
passes. The identical cells each have an active gas volume of about 30\,cm$^3$.  They were machined from copper and are equipped with anodes of 13-$\mu$m-diameter
gold-plated tungsten wire.  Methane at just over one-atmosphere pressure is continuously flushed through both cells.  Methane offers adequate gas gain for detecting
positrons and is quite insensitive to $\gamma$ radiation.  A Havar foil window, 3.7 cm in diameter and 1.5 $\mu$m thick, hermetically seals each gas cell on the side
facing the tape.  The performance of this type of detector is well known \cite{Ha78,Ko83,Ko97} and the particular detector employed in this measurement is the same one
used by us in many previous half-life measurements (\eg see Refs.\,\cite{Ia06, Pa11,Ia10, Pa12}).

Before and after our experiments, we check the performance of the gas counter using a $^{90}$Sr/$^{90}$Y $\beta$ source.  This source has been specially prepared on a sample
length of transport tape and is inserted into the gas counter in exactly the position that an on-line sample occupies.  We record the count rate as a function of the
applied counter bias voltage.  Initially, as the voltage increases the rate rises too since the increasing gas gain leads to more primary ionizing $\beta$ events producing a
large enough signal to be recorded.  However, at approximately 2600\,V a ``plateau'' is reached and the count rate remains nearly unchanged for the next 200-300\,V increase in the bias
voltage.  At higher voltages still, there is a second rapid rise in count rate as spurious pulses and electronic noise are increasingly recorded.  This
behavior is well understood \cite{Ha78} and clearly demonstrates that, when operated in the plateau region, such detectors have essentially 100\% efficiency with minimal
rate dependence.  During our $^{30}$S measurements, we operated the gas counter between 2650 and 2850\,V, well within the plateau region as determined with our source
measurements before and after the experiment.

\subsection{\label{position} Source positioning} 

Although the tape transport system is quite consistent in placing the collected source within $\pm$ 3\,mm of the center of the detector, it is a mechanical device and
occasionally larger deviations occur.  So, during the collection period of each cycle we recorded the number of ions passing through the scintillator en route to implantation;
then during the subsequent count period we recorded the number of positrons detected in the gas counter.  The ratio of the latter to the former is a sensitive measure of
whether the source was seriously misplaced in the counter.  We rejected the results from any cycle with an anomalous ({\it i.e.}~low) ratio.

\subsection{\label{analog} Data acquisition}

Data from this measurement were acquired with essentially the same analog electronics we have used in previous measurements; a detailed description
of the setup appears in Ref.\,\cite{Ia06}.  Briefly, the strongly amplified signals from the gas counter are fed to a low-threshold discriminator, the output of which is
split and sent to two fixed-width nonretriggering and nonextending gate generators, which introduce well known fixed dead times that are selected to be longer than any
up-stream dead-time contributions.  The outputs from the two gate generators are sent to two multi-channel scalers, each with 15,000 channels.  The time base for the
acquired spectra is defined by a Stanford Research System pulser, which is accurate to 0.01 ppm.

We set the widths of the two gate generators (dead times) to be different from one another and frequently changed them -- as well as the detector bias voltage and 
discriminator level -- from run to run as a check on any systematic effects that might arise from these adjustable parameters.  We monitored both gates continuously
during every run by recording coincidences between each gate and 4-ns-wide pulses from a precise 1-MHz pulse generator; this gave us a continuous measure of the
imposed dead times.  In this experiment we used dead times of 3 and 6\,$\mu$s in dead-time
channel A, and 6 and 4\,$\mu$s in channel B.

We also tested for any systematic count-rate dependence in our system.  Within a given run the initial $\beta$ counting rate was reasonably consistent from cycle to cycle.
However, we varied the $^{30}$S production rate so that some runs were recorded at relatively high rates, while others were at considerably lower rates.  Since $^{30}$S
was available in copious quantities, we covered initial rates from 7-20$\times10^3$ counts/s. 

For the first time, in this experiment we added another parallel data-acquisition branch to our analog system. The discriminator output signal was also sent to a multichannel
time-to-digital converter (TDC), which was used to record the arrival times of the $\beta$ signals.  Recorded in addition were the arrival times of the $^{30}$S ions detected
by the scintillator located just in front of the collection tape; and the logic signals that originated from the tape-transport control system, indicating the start of the
collect, move and detect periods.  Altogether, this information provided a decay spectrum and an independent measure of the time between collection and detection for each
cycle.  In this sense, it served as a parallel backup to our conventional equipment.

Beyond that, though, during analysis we could insert digitally into the time-stamp data stream from the TDC-based system, via software, any kind or amount of dead time we
chose.  We could then analyze the resulting decay histogram, making our usual dead-time correction but based on the dead time we inserted digitally.  The results in all cases were
found to agree closely with those from our conventional analysis; so they will not be referred to again.

\section{\label{results} Analysis and Results}

\subsection{\label{dataproc} Data processing}

Before analysis, we preselected the data based on two criteria.  First, a cycle was rejected if the total number of $\beta$ particles detected by the gas counter was less than 500,
which indicates that there had been little or no primary beam from the cyclotron during the collection period.  This happened infrequently, so only 1.4\% of the cycles
were eliminated by this criterion.  The second criterion we used to exclude cycles was the ratio of detected positrons to implanted sulfur ions; a low ratio indicates that the
implanted source was not centrally located in the gas counter for that cycle (see Sec.\,\ref{position}).  To be safe, for our final analysis of the data we set very tight limits on
this ratio, requiring it to be between 95 and 100\% of the maximum value obtained.  This eliminated 15.2\% of the total cycles.  A separate analysis with limits offset
at 91-98\% yielded a half-life that agreed with the one we obtained using our tight central limits to within their $\pm$0.02\% statistical uncertainties, thus confirming the validity
of the latter result.

The remaining data were then corrected cycle-by-cycle for dead-time losses. We used the conventional method, increasing the number of events in each channel in
the decay spectrum analytically for the known dominant dead-time; this method is described, for example, in Ref.\,\cite{Ko97}.  

In this measurement, we also added a further refinement: We corrected the data for the slight dependence of detection efficiency on counting rate.  This occurs because, even when the
detector is operated in the ``plateau'' region (see Sec.\,\ref{4pi}) there is a very slight increase in efficiency with bias voltage.  As the counting rate decreases during a decay
measurement, the current decreases across the series resistor that connects the anode wire to the bias supply, and the effective bias on the wire increases, thus increasing the
efficiency very slightly.  The details of our implementation of this correction will be described separately \cite{Ia18}.  Its impact on our final half-life turned out to be less than
0.02\%, a small but non-negligible amount. 

\begin{figure}[t]
\centering
\includegraphics[width=\columnwidth]{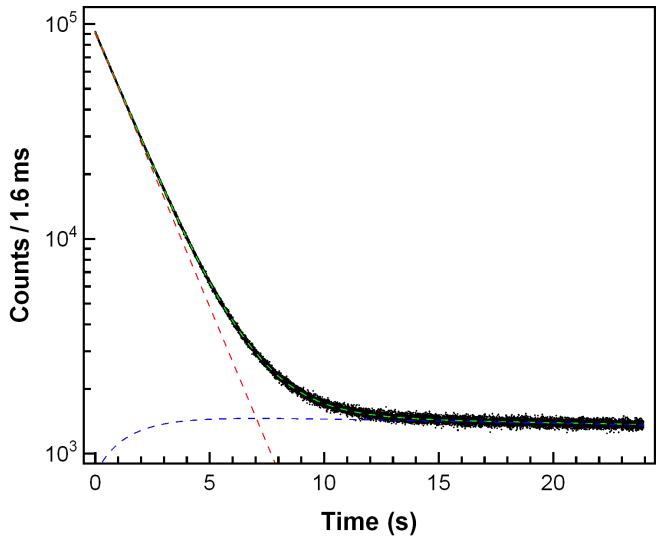}
\caption{\label{fig3} Measured time-decay spectrum for the total of all data obtained from the combined $\beta$ decays of $^{30}$S and its daughter $^{30}$P.  There are 15,000 data
points spanning 24\,s.  The dashed lines show the separate contributions calculated for the parent and daughter decays.  The constant background is off scale at 7.3 counts/1.6\,ms,
more than four orders of magnitude less than the initial source counting rate.  Note that counting began about 220\,ms after the beam was turned off.}
\end{figure}

The final decay spectrum for each run was then obtained from the sum of the dead-time- and efficiency-corrected decay spectra for all accepted cycles in that run.  To illustrate the
overall quality of the data, we present in Fig.\,\ref{fig3} the total time-decay spectrum obtained by combining all the runs.  Note that the daughter activity is entirely due to the
decay of the $^{30}$S parent, and at its maximum is a factor of $\sim$70 less than the initial $^{30}$S activity.  The constant background is another two orders of magnitude below that.

\subsection{\label{anresults} Half-life determination}

We fitted the data from each of the 33 runs separately, incorporating five components: $^{30}$S; its daughter $^{30}$P; impurities $^{29}$P and $^{27}$Si; and a constant background.
The half-life of $^{30}$P was fixed at its known value of 2.498(4)\,min \cite{Ba10} and the half-lives of $^{29}$P and $^{27}$Si were fixed at 4.140(16)\,s and 4.135(19)\,s \cite{Se08}
respectively.  As explained in Sec.\,\ref{purity}, the initial activities of $^{29}$P and $^{27}$Si relative to that of $^{30}$S were set at 0.050(15)\% and 0.015(15)\% respectively.
Finally, the background was fixed to the value obtained ($\sim$1 count/s) from a measurement in which all conditions were identical to those of a normal run except that the tape motion
was disabled.

\begin{figure}[t]
\centering
\includegraphics[width=\columnwidth]{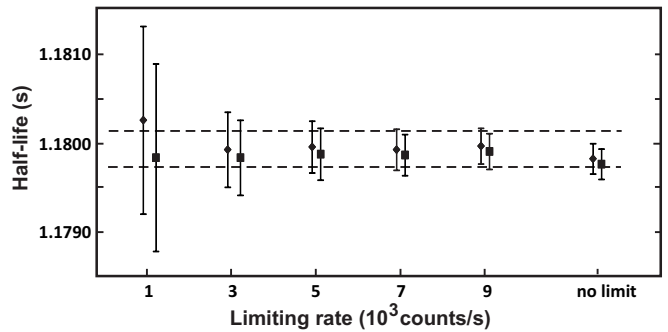}
\caption{\label{fig4} Test for any systematic dependence of our evaluated $^{30}$S half-life on the initial counting rate.  Each point represents the average half-life from all 33 runs as
obtained from only the portion of each decay spectrum that has a counting rate below a specified value.  The diamond-shaped points are take from dead-time channel A and the squares are from
channel B.  The dashed lines give the one-sigma uncertainty limits (statistical only) that correspond to the limiting rate of $9\times10^3$ counts/s.  Note that the results are strongly correlated
since the plotted points are derived from overlapping subsets of the same data.}
\end{figure}

Note that in our primary analysis we treat $^{30}$S and its daughter $^{30}$P as independent decay components.  Since all the $^{30}$P in our decay spectrum originates from the decay of
$^{30}$S, there is a well-defined connection between their intensities; and indeed, as explained in the Introduction, we have previously used such parent-daughter connections in fits
when the parent and daughter half-lives are so similar that the two components are not clearly separated in the decay spectrum \cite{Ia10,Ia06,Pa11}.  In the present case, however, the
half-life of $^{30}$P is more than 100 times longer than that of $^{30}$S, so there is no need to connect the two activities; and we deemed that by ignoring the connection we would
reduce our potential exposure to systematic uncertainties.  It is the result of this procedure that we quote as our final result.  However, we also performed a parallel analysis in
which the parent-daughter linkage was enforced in the fits.  By comparing the results from the two methods, we can test the efficacy of the latter, which has been the only method
available to us in previous measurements of $T_Z$\,=\,-1 superallowed $\beta$ emitters.

As explained in Sec.\,\ref{analog}, the initial counting rate associated with each run ranged from 7-20$\times10^3$ counts/s.  We therefore chose to analyze the runs in several different ways: First,
we accepted all the data from each run, regardless of its rate, and obtained a half-life value by averaging the fitted results from all 33 runs.  Next, we set a maximum limit of 9$\times10^3$ counts/s
and only analyzed that part of each run's decay spectrum for which the count rate had dropped below 9$\times10^3$ counts/s; if the run had a lower rate to start with, we used its full decay spectrum.
Finally we set lower rate limits -- at 7, 5, 3 and 1$\times10^3$ counts/s -- and repeated the analysis for each one.  The half-life results for the different rate limits are given in Fig.\,\ref{fig4};
the values for dead-time channels A and B are plotted separately, although they agree very closely with one another.

It is evident from Fig.\,\ref{fig4} that there is no significant change in the half-life result as the rate is substantially decreased.  In the past, we have routinely made measurements
at rates only up to 9$\times10^3$ counts/s and have found no rate-dependent effects in our data-acquisition system. Although the results in the figure indicate that the inclusion of higher-rate data makes
very little change, we have chosen to conservatively select the data set in which the rate is limited to 9$\times10^3$ counts/s as our standard, and it is the uncertainty limits on its
half-life value that appear as dashed lines in the figure.

\begin{figure}[t]
\centering
\includegraphics[width=\columnwidth]{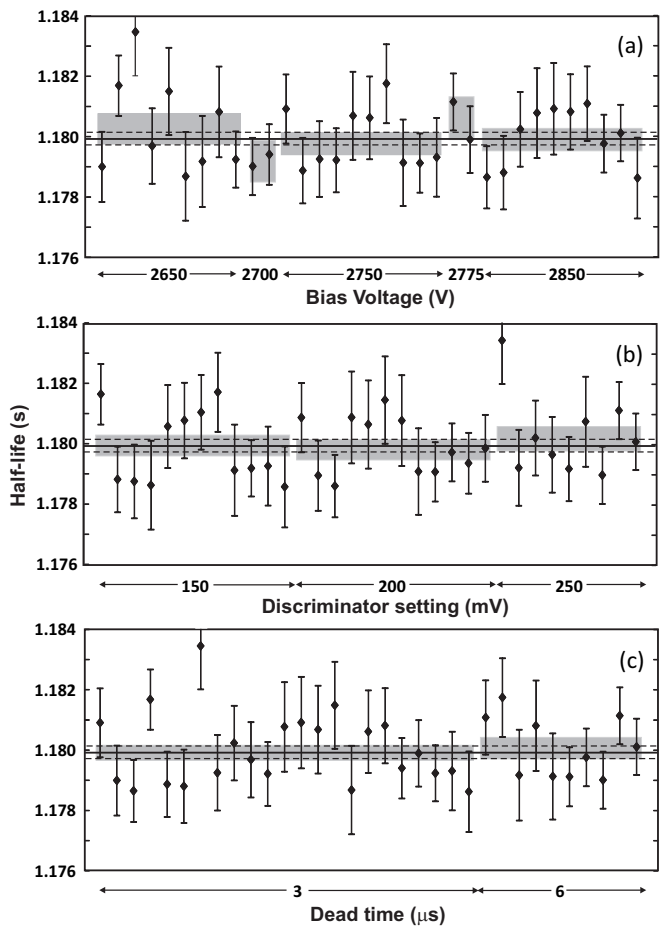}
\caption{\label{fig5} Test for any systematic bias in the $^{30}$S half-life measurement due to changes in three different detection parameters.  Shown are the results from dead-time
channel A for: (a) five detector biases; (b) three discriminator threshold settings; and (c) two imposed dead times.  Each graph includes half-life results from all 33 runs but the runs have
been grouped differently for each graph. In all cases, the grey bands represent the $\pm \sigma$ limits of the average half-life for a given condition. The average value for all the runs
appears as the solid horizontal line in each graph, with the corresponding dashed lines as the statistical uncertainty limits.  The value of the reduced $\chi^2$ for the 33-point average is 0.93.}
\end{figure}

\begin{figure}[t]
\centering
\includegraphics[width=\columnwidth]{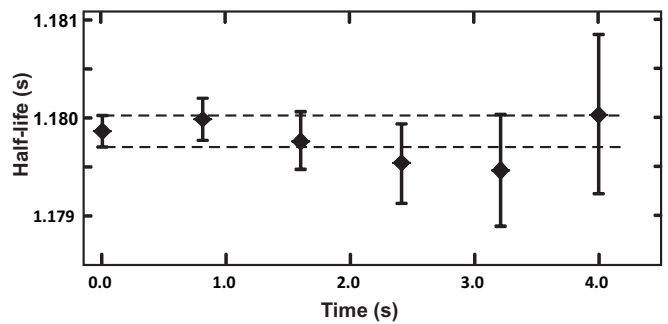}
\caption{\label{fig6} Test for any systematic bias in the $^{30}$S half-life results due to otherwise undetected short-lived impurities. Each point is the result of a separate
fit to the data from dead-time channel A; the abscissa for each point represents the time period at the beginning of the counting cycle for which the data were omitted from that
particular fit. The dashed lines correspond to the one-sigma uncertainty limits for the full data set.  Note that, unlike Fig.\,\ref{fig5}, this figure uses data without any
counting-rate limits applied (see text).}
\end{figure}

Since each run was executed with a different combination of detection settings -- bias voltage, discriminator threshold setting and dominant dead-time -- we could use the fitted half-lives of
$^{30}$S from each individual run to test for any systematic dependence on those settings.  Employing the data set that has been limited to a maximum rate of 9$\times10^3$ counts/s, we present the
results from dead-time channel A in Fig.\,\ref{fig5}.  The three parts of the figure each include all 33 runs, but the runs have been grouped differently in each part, according to the
parameter being investigated.  No statistically significant dependence on any of the three detection settings is apparent in the figure, and this observation applies equally well to
the results from dead-time channel B and from the data sets with other count-rate limits.

Our last systematic check was to test for any unanticipated short-lived impurities in the collected samples.  For this purpose we needed to return to the complete rate-unlimited data set
since, to be effective, this test requires that all decay spectra begin immediately after the source arrives at the counting location.  With this set, we removed data from the first
0.8\,s of the counting period and fitted the remaining data; then we removed an additional 0.8\,s of data and fitted again.  This procedure was continued until 4.0\,s of data -- 3.4
half-lives of $^{30}$S -- had been removed.  The result is given in Fig.\,\ref{fig6} where, as before, we have used the data from dead-time channel A; the results from channel B show
the same behavior.  Evidently the half-lives obtained in this way are consistent with one another within statistical uncertainties, ruling out any short-lived impurities.  This
test is also sensitive to any rate-dependent effects on the half-life measurement.  In agreement with the more sensitive test illustrated in Fig.\,\ref{fig4}, this test shows
no sign of rate dependence. 

As already explained in the context of Fig.\,\ref{fig4}, we take our final half-life result from the data set limited to 9$\times10^3$ counts/s, which is illustrated run-by-run in Fig.\,\ref{fig5}.
Our final average value for the $^{30}$S half-life, incorporating results from both dead-time channels is 1.17992(20)\,s with a reduced $\chi^2$ of 0.93.  At this stage, the
uncertainty is purely statistical.

With systematic instrumental effects eliminated as significant contributors to the half-life uncertainty, we must still take account of uncertainties arising from the known impurities in
the collected samples and from the half-life of $^{30}$P, the radioactive daughter of $^{30}$S, which is a growing presence in each decay spectrum.  The final error budget for the
$^{30}$S half-life is shown in Table\,\ref{ebudget}.  In this measurement, it is the uncertainty associated with the two weak impurities that is the largest contributor to the overall
uncertainty, although the counting statistics makes the second-largest contribution.  Our final result for the $^{30}$S half-life is 1.17992(34)\,s, in which statistical and
systematic uncertainties have been combined.

\begin{table}[b] 
\caption{\label{ebudget} Error budget for our $^{30}$S half-life measurement.}
\begin{ruledtabular}
\begin{tabular}{lc}
\multicolumn{1}{l}{Source}&
\multicolumn{1}{c}{Uncertainty (ms)} \\
\hline
statistics                                  & 0.20 \\
sample impurities ($^{29}$P,$^{27}$Si)      & 0.27 \\
$^{30}$P half-life                          & 0.016 \\
Total                                       & 0.34 \\ 
\\ 
$^{30}$S half-life result (s)  & 1.17992(34) \\
\end{tabular}
\end{ruledtabular}
\end{table}

\subsection{\label{comparison} Comparison with previous results}

\begin{figure}[t]
\centering
\includegraphics[width=0.99\columnwidth]{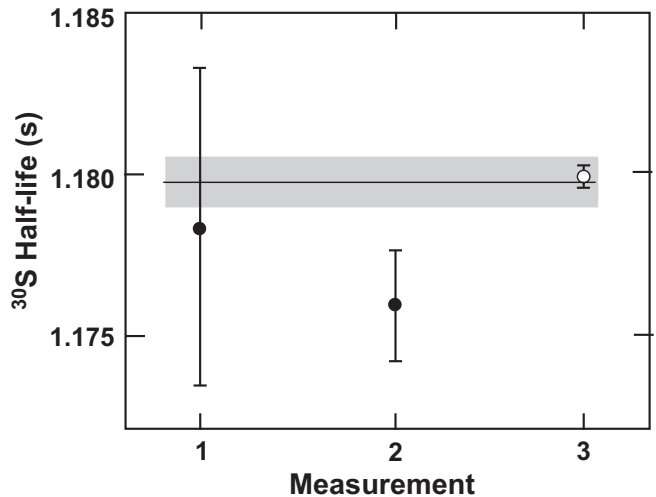}
\caption{\label{fig7} The present measurement (open circle) is compared with the only two previous measurements \cite{Wi80,So11} with sub-percent precision (solid circles).
The results are presented in chronological order from left to right.  The $\pm\sigma$ limits on the overall average value of 1.17976(77)\,s appear as the grey band.  Note
that the uncertainty of the average has been scaled by a factor of 2.3 as a result of the inconsistency between measurements 2 and 3 (see text).}
\end{figure}

Although several measurements of the $^{30}$S half-life have been reported previously, only two have small enough uncertainties to be relevant here.  These
are listed in the most recent review of $0^+$$\rightarrow$\,$0^+$ world data \cite{HT15}: 1.1783(48)\,s \cite{Wi80} and 1.1759(17)\,s \cite{So11}.    The first of these results was
published more than 35 years ago: The authors used a pneumatic shuttle to transport collected samples to a Ge(Li) detector, where 16 successive $\beta$-delayed $\gamma$-ray spectra
were recorded; the half-life was extracted from the decay of the 677-keV peak, the strongest one in the spectrum; this procedure uniquely isolated the decay of $^{30}$S.  The second,
more recent measurement, employed a pulsed beam and a ``close to 4$\pi$'' plastic scintillator to detect $\beta$ particles; both $^{30}$S and its daughter were produced
directly so the decay spectra, which included both activities, contained a relatively large contribution from the latter.  The results from these two previous measurements
appear in Fig.\,\ref{fig7}, together with our new result.  

With $\pm$0.029\% precision, our result is five times more precise than the most competitive previous measurement.  It agrees with one of the earlier results \cite{Wi80}, but differs
from the other \cite{So11} by more than two of the latter's standard deviations.  There is no obvious non-statistical explanation for this difference although we note that the ratio
of $^{30}$P-to-$^{30}$S activity in their spectra was ten times worse (\ie greater) than in our spectra.  The inclusion of the result from Ref.\,\cite{So11} in an overall average leads to a 
normalized $\chi^2$ of 2.7 and an average half-life for $^{30}$S of 1.17976(77)\,s, a result that includes an uncertainty scale factor (see Ref.\,\cite{HT15}) of 2.3.      

\subsection{\label{30P} Half-life of $^{30}$P }

In the midst of our measurement of the half-life of $^{30}$S, we changed the time sequencing of our collect-move-count cycles in order to focus on the decay of
its daughter, $^{30}$P.  Although the latter's half-life plays a minor role in the analysis of the $^{30}$S decay, we wanted to be sure that the accepted half-life
for $^{30}$P was not seriously incorrect.  With 30-s collect and 60-s detect periods we accumulated $\sim$2 million counts in 8 complete cycles.

These data were analyzed in the manner described in Sec.\,\ref{anresults} except that in this case the half-life of $^{30}$S was fixed at our measured value and the
half-life of $^{30}$P was determined from the fit.  The result we obtained for the half-life of $^{30}$P was 2.501(2)\,min, which is more precise than,
but completely consistent with, the literature value of 2.498(4)\,min \cite{Ba10}.

\subsection{\label{linkedtest} Test of analysis with parent-daughter link}

All the results quoted so far have been arrived at by our considering the $^{30}$S and $^{30}$P activities as being independent of one another.  We have already explained
that this is the optimum approach for parent and daughter activities with very different half-lives, but the same approach cannot be taken when the activities have similar
half-lives, as has been the case for the $T_Z$\,=\,-1 superallowed emitters we have studied previously.  As a test of the method we have been forced to use in those cases,
we have also analyzed the mass-30 data by fitting the decay spectra with the parent and daughter decays linked.  In doing so, we used our measured record of source deposition
as a function of time, and the calculated difference in detection efficiencies for the parent and daughter activities, just as we did in our previous measurements \cite{Ia06, Ia10, Pa11}.

The result we obtain for the $^{30}$S half-life that way is 1.17986(20)\,s, where the uncertainty is statistical only.  It agrees very closely with the result we quoted in
Sec.\,\ref{anresults} where we omitted the linkage between the parent and daughter activities, and the $\pm$0.017\% uncertainties are the same.  This convincingly validates
the method we have used in the past to determine precise half-lives for $T_Z$\,=\,-1 superallowed emitters.

\section{\label{concl} Conclusions}

We have measured the half-life of $^{30}$S to be 1.17992(34)\,s.  This is a factor of five more precise than the currently accepted value \cite{HT15} and, at $\pm$0.029\%,
is more than sufficient to allow the $^{30}$S superallowed $ft$ value to contribute to the evaluation of $V_{ud}$ once the branching ratio has been measured with sufficient
precision.  Unfortunately though, one of the previous half-life measurements \cite{So11} disagrees by two of its standard deviations from our more precise result, thus
degrading the precision attributable to the world-average value.  Using the methods employed in the periodic surveys of world data for superallowed decays (\eg see Ref.\,\cite{HT15}),
we obtain an average half-life value of 1.17976(77)\,s.  Although this scaled-up uncertainty ($\pm$0.065\%) is still an improvement over the previously accepted average, it
cries out for yet another measurement with sufficient precision (and accuracy) to resolve the disagreement now revealed between the two most-recent measurements.

Because the $^{30}$S and $^{30}$P half-lives differed by more than a factor of 100, we could analyze our data either by treating the parent and daughter activities as being
independent of one another, or by enforcing the linkage between them.  Thus, we could test the efficacy of the latter method, which is the one we have been forced to use
exclusively in previous half-life measurements of $T_Z$\,=\,-1 superallowed emitters, since in those cases the parent and daughter half-lives differed by merely a factor of
two.  The results from both methods were consistent well within their $\pm$0.017\% precision.

Finally, as a by-product of this measurement, we have measured the half-life of $^{30}$P to be 2.501(2)\,min.  This agrees with, but is a factor of two more precise than
the literature value.

\begin{acknowledgments}

This material is based upon work supported by the U.S. Department of Energy, Office of Science, Office of Nuclear Physics, under Award Number DE-FG03-93ER40773, and
by the Welch Foundation under Grant No.\,A-1397.

\end{acknowledgments}

\end{document}